\documentclass[12pt]{JHEP3}

\usepackage{epsfig}
\usepackage{citesort}

\def  \vtbvtd     {V_{tb} V_{td}^*}
\def  \vcbvcd     {V_{cb} V_{cd}^*}
\def  \vubvud     {V_{ub} V_{ud}^*}
\def  \vtbvts     {V_{tb} V_{ts}^*}
\def  \vcbvcs     {V_{cb} V_{cs}^*}
\def  \vubvus     {V_{ub} V_{us}^*}
\def  \lamu       {\lambda_u}

\def  \btopill    {B \to \pi \ell^+ \ell^-}
\def  \btorholl   {B \to \rho \ell^+ \ell^-}

\def  \btoxsll    {B \to X_s \ell^+ \ell^-}
\def  \btoxdll    {B \to X_d \ell^+ \ell^-}

\def  \btokksll   {B \to K(K^*) \ell^+ \ell^-}
\def  \btosll     {b \to s \ell^+ \ell^-}
\def  \btodll     {b \to d \ell^+ \ell^-}

\def  \bbtobdll   {\bar{b} \to \bar{d} \ell^+ \ell^-}

\def  \bdtollg    {B_d \to \ell^+ \ell^- \gamma}
\def  \bstoll     {B_s \to \ell^+ \ell^-} 
\def  \bstollg    {B_s \to \ell^+ \ell^- \gamma}
\def  \btoxdtt    {B \to X_d \tau^+ \tau^-}
\def  \bdtottg    {B_d \to \tau^+ \tau^- \gamma} 

\def  \btoxcenu   {B \to X_c  e \bar{\nu}_e }

\def  \csev       { C_7^{eff} }
\def  \cn         { C_9^{eff} } 
\def  \ct         { C_{10} }
\def  \cqo        { C_{Q_1} }
\def  \cqt        { C_{Q_2} }

\def  \bd         { \bar{d} }
\def  \bl         { \bar{\ell} }
\def  \mb         { m_b }

\def  \sh         { \hat{s} }
\def  \mlh        { \hat{m}_\ell }
\def  \mlhs       { \hat{m}_\ell^2 }

\def  \faco       { \sqrt{1 - \frac{4 \mlhs}{\sh}} } 

\def  \dgdsdz     { \frac{d \Gamma}{d \sh d\cos\theta}}

\def  \ev#1       { {\bf e}_#1 }

\def  \pl         { P_L }
\def  \pr         { P_R }
\def  \gf         { \gamma_5 }
\def  \gm         { \gamma_\mu }

\def  \gom        { \gamma^\mu }
\def  \sigmn      { \sigma_{\mu \nu} }

\def  \qon        { q^\nu }
\def  \qs         {q^2}

\def \epsmualbesig  {\varepsilon_{\mu\alpha\beta\sigma} }
\def \ppp        {p_+}
\def \ppm        {p_-}

\def \facmatrix {| \frac{\alpha^{3/2} G_F}{\sqrt{2 \pi}} V_{tb}
                 V_{td}^*|^2 } 

\def \modsq#1   {|#1|^2}
\def \rea#1#2   {Re(#1^* #2)}
\def \facdr     {| \frac{\alpha^{3/2} G_F}{2 \sqrt{2 \pi}} V_{tb}
                 V_{td}^*|^2 } 
\def \ctencqtwo {\left(C_{10} + \frac{m_{B_s}^2}{2 m_\ell m_b}
C_{Q_2}\right)} 
\def \drcqone   {\left(\frac{m_{B_s}^2}{2 m_\ell m_b} C_{Q_1}\right)}
\def \lnzh      {{\rm ln}(\hat{z})}

\def \psq       {p^2}
\def \p#1qsq    {(p_#1 q)^2}
\def \pqsq      {(p q)^2}
\def \ppqsq     {(p_+ q)^2}
\def \pmqsq     {(p_- q)^2}

\def \mlesq     {m_\ell^2 } 
\def \mlef      {m_\ell^4 }

\def \mbd       {m_{B_d}}
\def \mbdsq     {m_{B_d}^2}

\def \zh        {\hat{z}} 

\def \beq       {\begin{equation}}
\def \eeq       {\end{equation}}
\def \beqa      {\begin{eqnarray}}
\def \eeqa      {\end{eqnarray}}
\def \bfig      {\begin{figure}}
\def \efig      {\end{figure}}
\def \bcen      {\begin{center}}
\def \ecen      {\end{center}}

\def \ie        {{\it i.e. }}
\def \eg        {{\it e.g. }}
\def \etal      {{\it et. al. }}

\title{CP asymmetries in dileptonic decays of B-meson in MSSM}    

\author{S. Rai Choudhury 
and 
Naveen Gaur\footnote{URL : http://physics.du.ac.in/\~{}naveen}   \\
Department of Physics \& Astrophysics, \\
University of Delhi, \\
Delhi - 110 007, India. \\
E-mail : \email{src@ducos.ernet.in} ,\email{naveen@physics.du.ac.in}
}

\abstract{Scalar interactions (in effective Hamiltonian) can give
significant variation of  various experimental observables, as
comparted to their respective Standard Model values, in dileptonic
decays of B-meson. Also the quark level transition $\btodll$ can be
useful to test CP violation. Here we will do comparative study of CP
violation in two independent processes, which have the same quark
level transition ($\btodll$),  $\btoxdll$ (the inclusive decay mode)
and the exclusive channel $\bdtollg$ (radiative dileptonic decay
mode). We will mainly focus on the comparative study of scalar
interactions on the CP asymmetries in these two different channels.  
}

\preprint{\hepph{0207xxx}}

\begin{document}

\section{\label{section:1} Introduction}

The Flavor changing neutral current (FCNC) $b \to s(d)$ transition can
be a very useful probe of the weak interaction sector of SM because
this transition is forbidden in the tree approximation and goes
through a loop which is second order in weak interaction. In SM this
transition occurs through a intermediate $t, c$ or $u$ quark. Among
the processes having quark level $b \to s(d) $ transition, the ones 
having leptons in final state are more interesting because they are 
relative clean. The pure leptonic and semi-leptonic decays can also be
useful because they, over and above the branching ratio, can give us
many other experimentally measurable observable associated with pair
of final state leptons like lepton pair forward backward asymmetry (FB
asymmetry) and  the three polarization asymmetries \footnote{the three
polarization asymmetries are longitudinal, transverse and normal, in
pure leptonic mode, like $\bstoll$ there can only be one polarization
asymmetry which is  longitudinal because the kinematics of this mode
allows only one independent momenta}. These decays thus can be very
useful in testing the structure of effective Hamiltonian and can also
be used to test new physics beyond SM. One can also look at CP
violation in these transitions. If we look at $\btosll$, the
transitions involving intermediate $t$, $c$ and $u$ quarks enter with
CKM factors $\vtbvts$, $\vcbvcs$ and  $\vubvus$ respectively. Using
the Wolfenstein's parameterization of CKM matrix
\cite{Wolfenstein:1983yz} we can see that : 
$\vtbvts \ \sim \  \lambda^2$, $\vcbvcs \ \sim \ \lambda^2$ and
$\vubvus \ \sim \ \lambda^4$ where $\lambda \ = \ \sin\theta_C \ \cong
\ 0.22$. So we can see that $\vubvus$ can be neglected as compared to 
the other two. The unitarity relation for CKM factors hence reduces
to $ \vtbvts + \vcbvcs \ \approx \ 0$. So effectively we can remove
one in the favor of other and hence we are left with only one
overall CKM factor. Its phase will not show up in the transition rate
and hence CP-violation would not show up.

\par But the situation for $b \to d$ transition is different. Here the
contributions of intermediate $t$, $c$ and $u$ quarks are respectively
$\vtbvtd$ , $\vcbvcd$ and $\vubvud$ and all of these are of order
$\lambda^3$ and in general all three of them can have different
phase and hence the $\btodll$ transition rate would be sensitive to
CP-violating phases . This was studied in the case of inclusive
\cite{RaiChoudhury:1997a,Kruger:1996dt} channel and exclusive channel
\cite{Kruger:1997jk} within SM. Lately the scalar (and pseudoscalar)
interactions (in  effective Hamiltonian) have attracted lot of
interest in various purely leptonic
\cite{Skiba:1993mg,Choudhury:1999ze} and semi-leptonic decays like 
$\btopill, \btorholl$ \cite{Choudhury:2002ab,Demir:2002a} ,
$\btoxsll$
\cite{Dai:1997vg,Bobeth:2001jm,RaiChoudhury:1999qb,Grossman:1997qj}
$\btokksll$ \cite{Aliev:2000jx}, $\bdtollg$
\cite{Iltan:2000iw,RaiChoudhury:2002i}. The effects of the scalars on
CP asymmetries in the exclusive decays $\btopill$ and $\btorholl$ was
discussed in our earlier work \cite{Choudhury:2002ab}. But as
emphasized in some works \cite{RaiChoudhury:2002i,Iltan:2000iw} the
radiative dileptonic decay mode $\bstollg$ is also very sensitive to
the scalar interactions. This present work is a comparative study of
the CP asymmetries in inclusive dileptonic decay $\btoxdll$ and
exclusive decay $\bdtollg$. We will mainly focus on the effects of the
scalar interactions on the CP asymmetries of these two channels.    

\par The simplest and one of the most favourite extension of the SM
has been Minimal Supersymmetric extention of the SM (MSSM). In MSSM
there are five scalars (Higgs) as compared to one in SM. The
importance of these scalars also called as Neutral Higgs Bosons (NHBs)
have been extensively discussed  
\cite{Iltan:2000iw,Choudhury:2002ab,Skiba:1993mg,Dai:1997vg,Choudhury:1999ze,Bobeth:2001jm,RaiChoudhury:1999qb,RaiChoudhury:2002i,Demir:2002a}
in literature and we will use MSSM for our comparative study of CP
asymmetries. As known that in MSSM we have to include some additional
operators , over and above the usual SM operators in effective
Hamiltonian. These operators arise in MSSM because of the NHBs and the
coefficients (Wilson coefficients ) for with these operators  are
proportional to $m_\ell ~m_b ~tan^3\beta$, for large $\tan\beta$ which
means that the $\tau$ lepton processes would be affected most with a
much lesser effect for the ones with $\mu$. Here in our work we will
be going to take the final state leptons to be $\tau$. Although in SM
the Branching ratios of both $\btoxdtt$ ($\sim 10^{-8}$) and
$\bdtottg$ ($\sim 10^{-10}$) is very low  but it still might be
possible to observe it in future \eg in LHC-B where more than
$10^{11}$, $B_d$ mesons are expected to be produced. Also in MSSM
these branching ratios can be enhanced by an order in certain allowed
region of MSSM parameter space \footnote{in fact for radiative
dileptonic decay there can be a enhancement by two orders as we have
shown earlier for $\bstollg$ \cite{RaiChoudhury:2002i} }.     

\par The paper is organized as follows : In section \ref{section:2} we
will discuss the effective Hamiltonian for $\btodll$. In section
\ref{section:3} we will discuss CP violation in the inclusive decay
mode $\btoxdll$. In section \ref{section:4} we will discuss the
exclusive dileptonic decay mode $\bdtollg$ and finally in section
\ref{section:5} we will discuss our results and conclusions.   


\section{\label{section:2} The Effective Hamiltonian} 

The effective Hamiltonian for the decay $\btodll$ can be written as
\cite{Dai:1997vg} :
\beqa
{\cal H}_{eff} 
   &=& \frac{4 G_F \alpha }{\sqrt{2} \pi} \vtbvtd
       \Bigg[ \sum_{i = 1}^{10} C_i O_i 
          ~+~ \sum_{i=1}^{10} C_{Q_i} Q_i 
      ~-~ \lamu \left\{ ~C_1 [ O_1^u - O_1 ] 
    \right.  \nonumber  \\
   &&  \left. +~ C_2 [ O_2^u - O_2 ] ~
         \right\} 
       \Bigg] 
\label{sec2:eq:1}
\eeqa
where we have used the unitarity of the CKM matrix $\vtbvtd ~+~ \vubvud
\approx - \vcbvcd$, and $\lamu ~=~ \vubvud /\vtbvtd$. Here $O_1$ and
$O_2$ are the current current operators, $O_3, \dots, O_6$ are called
QCD penguin operators and $O_9$ and $O_{10}$ are semileptonic
electroweak penguin operators \cite{Ali:1997vt}\footnote{the only
difference being that s quark is replaced by d quark}. The new
operators $Q_i \rm (i = 1,\dots,10)$ arises due to NHB exchange
diagrams \cite{Skiba:1993mg,Dai:1997vg}. In this work we will use the
Wolfenstein parameterisation \cite{Wolfenstein:1983yz} of CKM matrix
with four real  parameters $\lambda, A, \rho ~{\rm and}~ \eta$ where
$\eta$ is the measure of CP violation. In terms of these parameters we
can write $\lamu$ as :    
\beq
\lamu ~=~ \frac{\rho (1 - \rho) - \eta^2}{(1 - \rho)^2 + \eta^2} 
~-~ i \frac{\eta}{(1 - \rho)^2 + \eta^2} ~+~ O(\lambda^2)
\label{sec2:eq:2}
\eeq
For inclusive decay we will also make use of :
\beq
\frac{|\vtbvtd|^2}{|V_{cb}|^2} ~=~  \lambda^2 [ (1 - \rho)^2 + \eta^2]
+ O(\lambda^4)  
\label{sec2:eq:3}
\eeq
The additional operators $O^u_{1,2}$ are :
\beqa
O^u_1 &=&  (\bd_\alpha \gm \pl u_\beta) ~(\bar{u}_\beta \gom \pl
b_\alpha)    \nonumber \\
O^u_2 &=&  (\bd_\alpha \gm \pl u_\alpha) ~(\bar{u}_\beta \gom \pl
b_\beta)    
\label{sec2:eq:4}
\eeqa
The resulting QCD corrected matrix element relevant to us can be
written as :
\beqa
{\cal M} 
 &=& \frac{G_F \alpha}{\sqrt{2} \pi} \vtbvtd 
   \left\{ ~ - 2 ~\csev ~\frac{\mb}{\qs} ~( \bd i \sigmn \qon \pr b )~
         ( \bl \gom \ell) ~+~ \cn ~ (\bd \gm \pl b)~ (\bl \gom \ell) 
   \right.      \nonumber  \\
 && \left.
      ~+~ \ct ~(\bd \gm \pl b )~ (\bl \gom \ell)  
    + ~\cqo ~(\bd \pr b)~(\bl \ell) 
    ~+~ \cqt ~(\bd \pr b)~ (\bl \gf \ell) ~  \right\}
\label{sec2:eq:5}
\eeqa
where q is the momentum transfer and $P_{L,R} ~=~ (1 \mp \gf)/2$ and
where we have neglected mass of d quark. The Wilson coefficients
$\csev$ and $\ct$ are given in many works 
\cite{Dai:1997vg,Cho:1996we,Grinstein:1989me} and the other Wilsons
$\cqo$ and $\cqt$ are given in
\cite{Choudhury:1999ze,Dai:1997vg}. The definition of $\cn$ is 
\cite{RaiChoudhury:1997a,Kruger:1996dt} :  
\beq
\cn = \xi_1 + \lambda_u ~ \xi_2
\label{sec2:eq:6}
\eeq
with 
\beqa
\xi_1 
  &=& 
    C_9 + g(\hat{m}_c, \sh) ( 3 C_1 + C_2 + 3 C_3 + C_4 + 3 C_5 +
    C_6) - {1 \over 2} g(\hat{m}_d, \sh) ( C_3 + 3 C_4 )  \nonumber \\
  && - {1 \over 2} g(\hat{m}_b, \sh) ( 4 C_3 + 4 C_4 + 3 C_5 + C_6) 
   ~+~ {2 \over 9} (3 C_3 + C_4 + 3 C_5 + C_6) 
\label{sec2:eq:7}            \\
\xi_2 
  &=& 
   [ g(\hat{m}_c, \sh) - g(\hat{m}_c, \sh) ] (3 C_1 + C_2) 
\eeqa
where 
\beqa
g(\hat m_i,\hat s) 
&=& 
  - {8 \over 9} ~ln(\hat m_i) ~+ ~{8 \over 27} ~+ ~{4 \over 9}~ y_i -
  ~{2 \over 9}~ (2 + y_i) ~\sqrt{|1 - y_i|}     \nonumber \\
&& 
  \times 
   \left\{ 
      \Theta(1 - y_i) 
         \left(ln\left(
                    \frac{1 + \sqrt{1 - y_i}}{1 - \sqrt{1 - y_i}}
                 \right) - i \pi 
         \right)
                    + \Theta(1 - y_i) ~2 ~arctan\frac{1}{\sqrt{y_i - 1}}    
   \right\} 
\label{sec2:eq:8}
\eeqa
with $y_i \equiv 4 \hat m_i^2/\hat s$ \footnote{hat over the masses
and momenta indicates that these are scaled quantities scaled by
$m_b$} . We will incorporate the long-distance contributions due to
charm quark resonances, \ie $c \bar{c}$ intermediate states, by using
the substitution
\cite{Dai:1997vg,RaiChoudhury:1999qb,Kruger:1997jk,Kruger:1996cv,Long-Distance}
:    
\beq
g(\hat{m}_c, \sh) ~\rightarrow ~g(\hat{m}_c, \sh) - 
\frac{3 \pi}{\alpha^2 }
\sum_{V = J/\psi,\psi',..}
\frac{M_V Br(V \to l^+ l^-)\Gamma_{total}^V}{(s - M_V^2)
+ i \Gamma_{total}^V M_V}   
\label{sec2:eq:9}
\eeq
we are now equipped with the effective Hamiltonian and the matrix
element and we proceed to calculate the CP asymmetries in next two
sections.   
 

\section{\label{section:3} Inclusive decay mode $\btoxdll$} 


 \subsection{\label{sec3:1} Decay rate and FB asymmetry}

The decay width as a function of invariant mass of lepton pair is
given by \cite{Dai:1997vg}: 
\beq
\frac{d \Gamma}{d \sh} 
  =  \frac{G_F^2 m_b^5}{768 \pi^5} \alpha^2 |\vtbvtd|^2 (1 - \sh)^2
\sqrt{1 - \frac{4 \mlhs}{\sh}} \Sigma_{\btoxdll}
\label{sec3:eq:1}
\eeq
where
\beqa
\Sigma_{\btoxdll} 
 &=& 
  4 |\csev|^2 \left(1 + \frac{2 \mlhs}{\sh}\right) \left(1 +
        \frac{2}{\sh} \right) 
  + |\cn|^2 \left(1 + \frac{2 \mlhs}{\sh} \right) (1 + 2 \sh)
   \nonumber  \\
 && + |\ct|^2 \left(1 - 8 \mlhs + 2 \sh + \frac{2 \mlhs}{\sh} \right) 
    + 12 Re(\csev \cn) \left(1 + \frac{2 \mlhs}{\sh} \right) 
   \nonumber \\
 && + {3 \over 2} |\cqo|^2 (\sh - 4 \mlhs)  + {3 \over 2} |\cqt|^2 \sh
    + 6 Re(\ct \cqt) \mlh
\label{sec3:eq:2}
\eeqa
To remove the uncertainties in the value of $m_b$ we normalize the
above decay rate to the charged current decay rate :
\beq
\Gamma(B \to X_c \ell \nu) = \frac{G_F^2 m_b^5}{192 \pi^3} |V_{cb}|^2 
f(\hat{m}_c) k(\hat{m}_c) 
\label{sec3:eq:3}
\eeq
where $f(\hat{m}_c)$ is the phase space factor and $k(\hat{m}_c)$ is
the QCD corrections to the semi-leptonic decay rate, these factors are
given in appendix. The differential branching ratio hence becomes :
\beq
\frac{d Br(\btoxdll)}{d \sh} = \frac{\alpha^2}{4 \pi^2}
       \frac{ |\vtbvtd|^2 }{ |V_{cb}|^2 }
\frac{Br(\btoxcenu)}{f(\hat{m}_c) \kappa(\hat{m}_c) }
\Sigma_{\btoxdll} 
\label{sec3:eq:4}
\eeq

\vspace*{1cm}
\FIGURE[h]
{
\epsfig{file=incl_bdr.eps,width=0.9\textwidth}
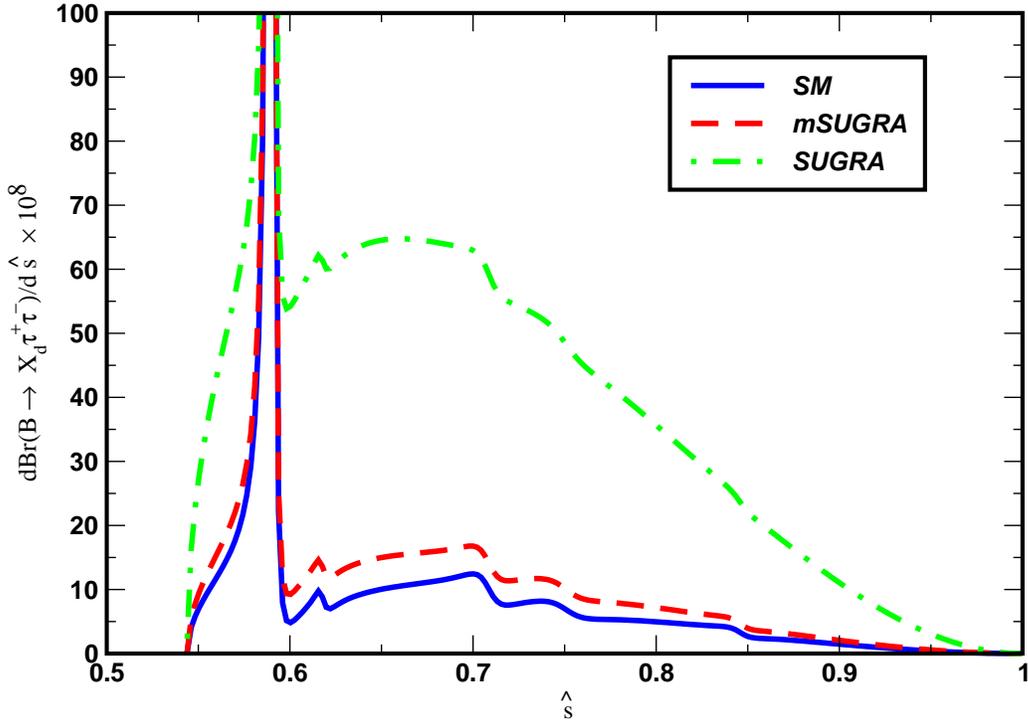
\caption{Branching ratio of $\btoxdtt$ with invariant mass of
dileptons. All the parameters of mSUGRA and SUGRA are given in
appendix \ref{appendix:1} }  
\label{fig:1}
}

As has been earlier on also mentioned that FB asymmetry is also very
sensitive to the new physics. For completeness we give the expression
of FB asymmetry also. The definition of the FB asymmetry is :
\beq
A_{FB} ~=~  
   \frac{ \int_0^1 d\cos\theta  \dgdsdz - \int_{-1}^0 d\cos\theta
       \dgdsdz }{\int_0^1 d\cos\theta  \dgdsdz + \int_{-1}^0
       d\cos\theta \dgdsdz }
\label{sec3:eq:5}
\eeq
where $\theta$ is the angle between the momentum of B-meson and
$\ell^-$ in the CM frame of dileptons. The analytical expression of
the FB asymmetry is :
\beq
A_{FB}(\sh) ~=~ \frac{6 \left(1 - \frac{4
\mlhs}{\sh}\right)}{\Sigma_{\btoxdll}} 
  Re \Bigg[ 2 \csev \ct + \cn \ct \sh + 2 \csev \cqt \mlh + \cn \cqo
     \mlh  \Bigg] 
\label{sec3:eq:6}
\eeq

\FIGURE[ht]
{
\epsfig{file=incl_bfb.eps,width=0.9\textwidth}
\caption{FB asymmetry of $\btoxdtt$ with invariant mass of
dileptons }  
\label{fig:2}
}


\subsection{\label{sec3:2} CP asymmetries} 

Next we define the CP violating partial width asymmetry as :
\beq
A_{CP}(\sh) ~=~ \frac{\frac{d \Gamma}{d\sh} - \frac{d
\bar{\Gamma}}{d\sh}}{\frac{d \Gamma}{d\sh} + \frac{d
\bar{\Gamma}}{d\sh}} 
\label{sec3:eq:7}
\eeq
where
\beq
\frac{d\Gamma}{d\sh} ~=~ \frac{d\Gamma(\btodll)}{d\sh} ~~,~~
\frac{d\bar{\Gamma}}{d\sh} ~=~ \frac{d\Gamma(\bbtobdll)}{d\sh}
\label{sec3:eq:8}
\eeq
In going from $\Gamma$ to $\bar{\Gamma}$ the only change would be in
the term having $\cn$ in the matrix element. The definition of $\cn$
is given in eqn(\ref{sec2:eq:6}). Now to find $\bar{\Gamma}$ the
definition of $\cn$ changes to :
\beq
\cn ~=~ \xi_1 + \lambda_u^* \xi_2
\label{sec3:eq:9}
\eeq
one can easily calculate the expression of CP-violating partial width
asymmetry from the expression of decay width eqn.(\ref{sec3:eq:1}) ,
the expression of CP-violating partial width asymmetry is :
\beq
A_{CP}(\sh) = \frac{- 2 Im \lamu
\bigtriangleup_{\btoxdll}}{\Sigma_{\btoxdll} + 2 Im \lamu
\bigtriangleup_{\btoxdll}} 
\label{sec3:eq:10}
\eeq
where $\Sigma_{\btoxdll}$ is given in eqn.(\ref{sec3:eq:2}) and
$\bigtriangleup_{\btoxdll}$ 
is :
\beq
\bigtriangleup_{\btoxdll} ~=~  Im(\xi_1^* \xi_2) \left(1 + \frac{2
\mlhs}{\sh} \right)(1 + 2 \sh) + 6 Im(\csev \xi_2) \left(1 + \frac{2
\mlhs}{\sh} \right)  
\label{sec3:eq:11}
\eeq

\vskip 1cm
\FIGURE[h]
{
\epsfig{file=incl_cpdr.eps,width=0.9\textwidth}
\caption{CP violating asymmetry $A_{CP}$ in $\btoxdtt$ with invariant
mass of dileptons.  }  
\label{fig:3}
}

As argued in many earlier works
\cite{RaiChoudhury:1997a,Choudhury:2002ab,Kruger:1997jk} that by
measuring the FB asymmetries of $B$ and $\bar{B}$ also one can observe
the CP violating phase of the CKM matrix. 

\par While discussing the CP violation through the FB asymmetries it
is important to fix up the sign convention. The reason for this is
that there are generally two conventions available in litreature 
regarding this sign. One is followed by Kr\"{u}ger and Sehgal
\cite{Kruger:1997jk} where the {\sl difference} of FB asymmetries of
$B$ and $\bar{B}$ was taken as the measure of CP violation. The other 
convention is where the {\sl sum} of FB asymmetries of $B$ and
$\bar{B}$ is taken to be the extent of CP violation
\cite{RaiChoudhury:1997a}.  Actually both these conventions are same,
the reason for this is that sign of FB asymmetry for $B$ and $\bar{B}$
are different. In fact in the limit of strict CP conservation :
\beq
A_{FB}(\bar{B}) ~=~ - ~A_{FB}(B) 
\label{sec3:eq:12}
\eeq
We can easily understand this because CP conjugation not only requires
exchange $b \leftrightarrow \bar{b}$ but also $\ell^- \leftrightarrow
\ell^+$. Since the two dileptons are emmited back to back in dilepton 
CM frame, the asymmetry defined in terms of direction of $\ell^-$ (for
both $B$ and $\bar{B}$) changes sign under CP transformation
\footnote{Kr\"{u}ger \& Sehgal \cite{Kruger:1997jk} haven't considered
this sign change or in other words for $B$ they calculate FB asymmetry
wrt $\ell^-$ but for $\bar{B}$ they calculate FB asymmetry wrt
$\ell^+$} . Any deviation from eqn.(\ref{sec3:eq:12}) will give us
another measure of CP violation. We for this define a CP violating
parameter in FB asymmetry as :
\beq
\delta_{FB} = A_{FB}(B) + A_{FB}(\bar{B})
\label{sec3:eq:13}
\eeq
Using the expressionf of the FB asymmetry eqn.(\ref{sec3:eq:6}) we can
get :
\beq
\delta_{FB} = \frac{2 Im \lamu
  \Bigg[ - Im \xi_2 ( \ct \sh + \cqo \mlh) \Sigma_{\btoxdll} + 2
\bigtriangleup_{\btoxdll} N_1 \Bigg] 
}{\Sigma_{\btoxdll} ( \Sigma_{\btoxdll}
+ 4 Im \lamu \bigtriangleup_{\btoxdll} )}  
\label{sec3:eq:14}
\eeq
with 
\beqa
N_1 
  &=& 
    2 \csev \ct + ( Re \xi_1 + Re \lamu Re \xi_2 
     - Im \lamu Im \xi_2 ) ( \ct \sh + \cqo \mlh)  \nonumber \\ 
  && + 2 \csev \cqt \mlh  
\label{sec3:eq:15}
\eeqa
and $\Sigma_{\btoxdll}$ is given in eqn.(\ref{sec3:eq:2}) 

\FIGURE[ht]
{
\epsfig{file=incl_cpfb.eps,width=0.9\textwidth}
\caption{CP violating asymmetry $\delta_{CP}$ in $\btoxdtt$ with
invariant mass of dileptons }  
\label{fig:4}
}


\section{\label{section:4} Exclusive decay mode $\bdtollg$}


\subsection{\label{sec4:1} Decay rate and FB asymmetry } 

The procedure for calculation of the decay rate of $\bdtollg$ is
exactly same as that of $\bstollg$
\cite{Iltan:2000iw,RaiChoudhury:2002i} with the replacement $s
\rightarrow d$ . As explained earlier
\cite{Iltan:2000iw,RaiChoudhury:2002i} the exclusive $\bdtollg$ decay
is induced by the inclusive $\btodll$ one. So, we have to start with
QCD corrected effective Hamiltonian for related quark level process
$\btodll$ given in eqn.(\ref{sec2:eq:1})   

\par In order to obtain the matrix element for $\bdtollg$ decay, a 
photon line should be hooked to any of the charged internal or
external lines. As has been pointed out before \cite{Aliev:1997ud}, 
contributions coming from hooking a photon line from any charged
internal line will be suppressed by a factor of $m_b/M_W^2$, and hence
we neglect them in our further analysis. When photon is attached to
the initial quark lines the corresponding matrix element is the {\sl
so called} {\tt  structure dependent} (SD) part of the amplitude which
can be written as :
\beqa
{\cal M}_{SD} = \frac{\alpha^{3/2} G_F}{\sqrt{2 \pi}} V_{tb} V_{td}^* 
           &&  \left\{ ~[ A ~\epsmualbesig {\epsilon^*}^\alpha p^\beta
                  q^\sigma ~+~ i B ~
                 (\epsilon_\mu^*(pq)-(\epsilon^*p)q_\mu)] ~ 
                 \bar{\ell}\gamma^\mu\ell
               \right.                                 \nonumber\\
           && + ~ \left.[C ~\epsmualbesig {\epsilon^*}^\alpha p^\beta
                   q^\sigma ~+~ i D ~
                   (\epsilon_\mu^*(pq)-(\epsilon^*p)q_\mu) ] 
                   \bar{\ell}\gamma^\mu\gamma_5\ell   
                 \right\}
\label{sec4:eq:1}
\eeqa 
where definition of form factors and A, B, C and D are given in
appendix (\ref{appendix:2}). In the defination of A and B (given in
eqn.(\ref{appen2:eq:1}) the value of $\cn$ is given by
eqn.(\ref{sec2:eq:6}).  We can see from eqn.(\ref{sec4:eq:1})   
that neutral scalar exchange parts do not contribute to the {\tt
structure dependent} part. 

\par When the photon is attached to the lepton lines using the
eqns.(\ref{appen2:eq:6},\ref{appen2:eq:7},\ref{appen2:eq:8}) 
and the cons\-ervation of vec\-tor cur\-rent we can get the
con\-tri\-bu\-tion to \- the Bremss\-trahlung part (called {\tt
internal Bremsstrahlung} IB) part as : 
\beqa
{\cal M}_{IB}
   &=&  \frac{\alpha^{3/2} G_F}{\sqrt{2\pi}} ~V_{tb} V_{td}^* ~
       i 2 ~m_{\ell} ~
        f_{B_d} ~
        \left\{ ~( C_{10} ~+~ \frac{m_{B_d}^2}{2 m_{\ell} m_b} \cqt) ~
           \bar{\ell} \left[ \frac{\not\epsilon \not P_{B_d}}{2 \ppp q}
              ~-~ \frac{\not P_{B_d}\not\epsilon}{2 \ppm q}
                      \right] \gamma_5\ell
        \right.                              \nonumber        \\
   &&  ~+~ \left. \frac{m_{B_d}^2}{2 m_{\ell} m_b} \cqo 
           \Bigg[ 2 m_\ell ( \frac{1}{2 \ppm q} ~+~ \frac{1}{2 \ppp q})~ 
                  \bar{\ell}\not\epsilon\ell
      \right.         \nonumber   \\
   && \left. +~ \bar{\ell} ~ ( \frac{\not\epsilon \not P_{B_d}}{2 \ppp q}
        ~-~ \frac{\not P_{B_d}\not\epsilon}{2 \ppm q}) ~ \ell ~
          \Bigg] ~
        \right\}.
\label{sec4:eq:2}
\eeqa
where $P_{B_d}$ and $f_{B_d}$ are the momentum and decay constant of
the $B_d$ meson. $\ppm$ and $\ppp$ are the four momental of $\ell^-$ and
$\ell^+$ respectively. 

\par The total matrix element for $\bdtollg$ is obtained as a sum of
${\cal M}_{SD}$ and ${\cal M}_{IB}$ terms :
\beq
{\cal M} ~=~ {\cal M}_{SD} ~+~ {\cal M}_{IB}
\label{sec4:eq:3}
\eeq
From above matrix element we can get the square of the
matrix element as,(with photon polarizations summed over) 
\beq
\sum_{\rm photon ~pol} |{\cal M}|^2 ~=~ |{\cal M}_{SD}|^2 ~+~ |{\cal 
M}_{IB}|^2 ~+~ 2 Re({\cal M}_{SD} {\cal M}_{IB}^*)  
\label{sec4:eq:4}
\eeq
with 
\beqa
|{\cal M}_{SD}|^2 &=& 4 ~ \facmatrix ~ 
             \left\{ 
            ~[ ~\modsq{A} + \modsq{B}~ ] ~ [ \psq ( \pmqsq + \ppqsq ) 
               + 2 m_\ell^2 \pqsq ]   
             \right.   \nonumber \\
         &&    \left.+~ [~ \modsq{C}  ~+~ \modsq{D}~ ] 
          ~ [ \psq ( \pmqsq +  \ppqsq ) - 2 m_\ell^2 \pqsq ]
              \right.           \nonumber \\
         && \left. +~ 2 ~Re(B^* C + A^*D)~ \psq ( \ppqsq - \pmqsq ) 
             \right\}    
\label{sec4:eq:5}             
\eeqa
\beqa
|{\cal M}_{IB}|^2  &=& 
   4 ~ \facmatrix ~ f_{B_d}^2 ~ \mlesq 
       \Bigg[
	  \ctencqtwo 
           \left\{ ~ 8 + 
              \frac{1}{\pmqsq } 
            \right.     \nonumber \\
 &&         \left. \times ( - 2 \mbdsq \mlesq - \mbdsq \psq + p^4 ~+~
                 2 p^2 ( \ppp q ) )
         .  ~+~  \frac{1}{(\ppm q)} ( 6 \psq + 4 (\ppp q) )
	   \right.            \nonumber  \\
 &&        \left. +~  \frac{1}{\ppqsq } 
           ( - 2 \mbdsq \mlesq - \mbdsq \psq + p^4 + 2 \psq (\ppm q) )
            +~  \frac{1}{(\ppp q)} ( 6 \psq + 4 (\ppm q) )
           \right.        \nonumber   \\
 &&        \left. +~ \frac{1}{(\ppm q) (\ppp q)} ( - 4 \mbdsq \mlesq + 2
           p^4 ) 
            ~ \right\}  ~+~  \drcqone 
         \left\{ 8  ~+~  \frac{1}{\pmqsq } 
         \right.         \nonumber  \\
 &&      \left. 
          \times ( 6 \mbdsq \mlesq + 8 \mlef - \mbdsq \psq - 8 \mlesq \psq 
            + p^4 - 8 \mlesq (\ppp q) + 2 \psq (\ppp q) ) 
         \right.          \nonumber \\
 &&      \left. 
             +~ \frac{1}{ (\ppm q) } (- 40 \mlesq + 6 \psq + 4 (\ppp q) )
             ~+~  \frac{1}{\ppqsq } ( 6 \mbdsq \mlesq + 8 \mlef - \mbdsq
           \psq 
         \right.         \nonumber  \\
 &&      \left.  
         - 8 \mlesq \psq + p^4 - 8 \mlesq (\ppm q) 
           +~ 2 \psq (\ppm q) )
             ~+~ \frac{1}{ (\ppp q) } ( - 40 \mlesq + 6
              \psq  + 
         \right.  \nonumber   \\
 &&      \left. 
        + 4 (\ppp q) ) ~+~ \frac{1}{(\ppm q) (\ppp q)} ( 4 \mbdsq \mlesq +
          16 \mlef  - 16 \mlesq p^2  + 2 p^4 ) 
         \right\}
	\Bigg]                                    
\label{sec4:eq:6}              
\eeqa
\beqa
2 Re( {\cal M}_{SD} {\cal M}_{IB}^* ) 
 &=&  
16 ~ \facmatrix  f_{B_d} ~ \mlesq 
       \Bigg[  \ctencqtwo 
               \left\{ - ~ Re(A)   
               \right.           \nonumber \\
 &&            \left. 
               \times \frac{( \ppm q + \ppp q)^3}{(\ppm q) (\ppp q)} 
                ~+~ Re(D) \frac{ (p q)^2 ( \ppm q ~-~ \ppp q ) }
                        {(\ppm q) (\ppp q)} 
               \right\}    \nonumber \\
 &&            + \drcqone
               \left\{ \frac{Re(B)}{(\ppm q)(\ppp q)} ( - (p q)^3
                - 2 (\ppm \ppp) (\ppp q)^2 
               \right.          \nonumber \\
 &&             \left.  - 2 (\ppm \ppp) (\ppm q)^2 
             + ~4 \mlesq (\ppm q)(\ppp q))  
                  ~ +~ Re(C)   \right.     \nonumber  \\
 &&           \left. 
              \times \frac{(p q)^2 ( \ppm q - \ppp q )}{(\ppm q) (\ppp
               q)}  
               \right\}
\label{sec4:eq:7}
\eeqa
The differential decay rate of $\bdtollg$ as a function of invariant mass
of lepton pair is given by:
\beqa
\frac{d \Gamma}{d\sh} 
   = ~\facdr ~\frac{\mbd^5}{16 (2 \pi)^3}~ (1 - \sh) ~ \faco
      ~ \Sigma_{\bdtollg}
\label{sec4:eq:8}
\eeqa
with $\Sigma_{\bdtollg}$ defined as
\beqa
\Sigma_{\bdtollg} &=& 
      {4 \over 3} ~\mbdsq ~( 1 - \sh)^2~ 
        [ ~ ( \modsq{A} + \modsq{B} )~ (2 \mlhs + \sh) 
          ~+~ ( \modsq{C} + \modsq{D} ) ( - 4 \mlhs + \sh) ~ ]
                                                     \nonumber \\
  &&  +~ \frac{64 f_{B_d}^2 \mlhs}{\mbd^2} \ctencqtwo^2 \frac{[~
        (1 - 4 \mlhs + \sh^2) \lnzh - 2 \sh \faco ] }{(1 - 
          \sh)^2  \faco}                            \nonumber \\ 
  &&  -~ \frac{64 ~f_{B_d}^2 \mlhs}{\mbd^2} ~ \drcqone^2 ~ 
         \left\{ \frac{( - 1 + 12 \mlhs - 16 \mlh^4 - \sh^2)
           \lnzh}{(1 - \sh)^2 \faco}      
         \right.          \nonumber     \\
  &&       \left.   ~+~ \frac{( -2 \sh - 8 \mlhs \sh + 4 \sh^2 )}{(1 -
           \sh)^2} \right\}
        ~+~ 32 ~ f_{B_d} \mlhs ~\ctencqtwo ~ Re(A) 
                 \nonumber  \\
  &&    \times   ~\frac{(-1 + \sh) \lnzh}{\faco}       
      -~ 32 ~ f_{B_d} \mlhs ~\drcqone ~Re(B)            \nonumber  \\
  &&  \times \frac{[ ~ (1 - 4 \mlhs + \sh) \lnzh ~-~ 2 \sh \faco
~]}{\faco}   
\label{sec4:eq:9}
\eeqa
where $\sh = \psq/\mbdsq ~,~ \mlhs = m_\ell^2/\mbdsq ~,~ \zh = \frac{1 + 
\faco}{1 - \faco}$ are dimensionless quantities. 

\FIGURE[ht]
{
\epsfig{file=bllg_bdr.eps,width=0.9\textwidth}
\caption{Branching ratio of $\bdtottg$ with invariant mass of
dileptons }   
\label{fig:5}
}

We can also calculate the FB asymmetry from use of
eqn.(\ref{sec3:eq:5}).  The analytical expression of FB asymmetry is :
\beqa
A_{FB} &=& 
      \Bigg[
        - 2 ~\mbdsq ~Re(A^* D + B^* C) ~(1 - \sh)^2 ~\sh ~ \faco 
                 \nonumber   \\
     && +~~ 32 ~f_{B_d} ~ \mlesq ~ \frac{(-1 + \sh)}{\faco} 
       Log\left(\frac{4 \hat m_\ell^2 }{\sh} \right) 
        \left\{\ctencqtwo Re(D) 
        \right.         \nonumber   \\
     && \left. + \drcqone Re(C)  \right\} 
      \Bigg]{\Bigg /}\Sigma_{\bdtollg}
\label{sec4:eq:10}
\eeqa

\FIGURE[ht]
{
\epsfig{file=bllg_bfb.eps,width=0.9\textwidth}
\caption{FB asymmetry of $\bdtollg$ with invariant mass of dileptons }
\label{fig:6}
}


\subsection{\label{sec4:2} CP asymmetries} 

One can also calculate the CP asymmetries as defined in
eqn.(\ref{sec3:eq:7}) and eqn.(\ref{sec3:eq:13}). The expression of CP
violating partial width asymmetry is :
\beq
A_{CP} = \frac{- 2 Im\lamu
\bigtriangleup_{\bdtollg}}{\Sigma_{\bdtollg} + 2 Im \lamu
\bigtriangleup_{\bdtollg}} 
\label{sec4:eq:11}
\eeq
with $\Sigma_{\bdtollg}$ given in eqn.(\ref{sec4:eq:9}) and expression
of $\bigtriangleup_{\bdtollg}$ is :
\beqa
\bigtriangleup_{\bdtollg} 
 &=& 
  \Bigg[ 
   \left\{ G_1(\psq) + F_1(\psq) \right\} Im(\xi_1^* \xi_2) 
          \nonumber \\ 
 &&     - \frac{2 \mb}{\psq} \left\{ G_1(\psq) G_2(\psq) + F_1(\psq)
      F_2(\psq) \right\} \ct Im(\xi_2) 
  \Bigg] \times T_1(\sh,\mlh)  \nonumber   \\
 &&  + ~ \ctencqtwo G_1(\psq) T_2(\sh,\mlh) \times Im \xi_2  
      \nonumber   \\ 
 && + \drcqone F_1(\psq) T_3(\sh,\mlh) \times Im \xi_2 
\label{sec4:eq:12}
\eeqa
with
\beqa
T_1(\sh,\mlh) &=& \frac{1}{\mbd^2} \frac{4 (1 - \sh)^2 (2 \mlhs +
\sh)}{3}   
\label{sec4:eq:13}    \\
T_2(\sh,\mlh) &=& 16 \frac{f_{B_d}}{\mbd^2} \frac{\mlhs (-1 + \sh)
ln(\zh)}{\faco}
\label{sec4:eq:14}    \\
T_3(\sh,\mlh) &=& - 16 \frac{f_{B_d}}{\mbd^2} \frac{[ (1 - 4 \mlhs +
\sh) ln(\zh) - 2 \sh \faco]}{\faco} 
\label{sec4:eq:15}
\eeqa

\FIGURE[ht]
{
\epsfig{file=bllg_cpdr.eps,width=0.9\textwidth}
\caption{CP violating asymmetry $A_{CP}$ in $\bdtottg$ with
invariant mass of dileptons }  
\label{fig:7}
}

Similarly we can calculate the second CP violating parameter
$\delta_{FB}$ as defined in eqn.(\ref{sec3:eq:13}). The expression of
$\delta_{FB}$ is : 
\beqa
\delta_{FB} = \frac{2 Im \lamu \times [- \Sigma_{\bdtollg} L_1 + 2
\bigtriangleup_{\bdtollg} L_2]}{\Sigma_{\bdtollg}(\Sigma_{\bdtollg} + 4
Im \lamu \bigtriangleup_{\bdtollg})} 
\label{sec4:eq:16}
\eeqa
with $\Sigma_{\bdtollg}$ and $\bigtriangleup_{\bdtollg}$ are given in
eqns.(\ref{sec4:eq:9}) and (\ref{sec4:eq:12}) respectively. $L_2$ is
just the numerator of the expression of FB asymmetry in
eqn.(\ref{sec4:eq:10}) and $L_1$ is given as :
\beqa
L_1 = - 2 (1 - \sh)^2 \sh \faco 
  \Bigg[ D G_1(\psq) Im(\xi_2) + C F_1(\psq) Im(\xi_2) \Bigg]  
\label{sec4:eq:17}
\eeqa

\FIGURE[ht]
{
\epsfig{file=bllg_cpfb.eps,width=0.9\textwidth}
\caption{CP violating asymmetry $\delta_{CP}$ in $\bdtollg$ with
invariant mass of dileptons }  
\label{fig:8}
}


\section{\label{section:5} Results and discussion }

We have performed the numerical analysis of all the asymmetries,
branching ratios and FB asymmetries whose analytical expressions are
given in previous sections. 

\par The MSSM that we are working with is the simplest (and having the
least number of parameters) SUSY model, but even this still has too
many of parameters to do any meaningful phenomenology with it. There
are many choices available to restrict this large parameter space. We
have opted for  Supergravity (SUGRA) model for our analysis. In this
model the universality of all the scalar masses and coupling constants
at the unification scale is assumed. So in minimal SUGRA (mSUGRA)
model we only have five parameters (in addition to SM parameters)
namely : $m$ the unified mass of all the scalars at GUT scale , $M$
the unified gaugino mass at GUT scale, $A$ the universal trilinear 
coupling at unification scale , $tan\beta$ the ratio of vacuum
expectation values of the two Higgs doublets and finally
$sgn(\mu)$. We have also considered another model where we have
relaxed the condition of the universality of the scalar masses at GUT
scale. This sort of model lately has been advocated in many works 
\cite{goto1,Choudhury:2002ab,Choudhury:1999ze,RaiChoudhury:1999qb,RaiChoudhury:2002i}
In this model we have
taken the squark sector and Higgs sector to have different unified
masses at GUT scale. So here we have another parameter which we have
taken to be the pseudoscalar Higgs mass \footnote{our choice of
parameters is given in Appendix \ref{appendix:1}}. About the sign of
convention of $\mu$, we are following the convention where $\mu$
enters the chargino mass matrix with +ve sign. In all of our numerical
analysis we have taken a 95\% CL bound  \cite{expbsg} 
$$ 2 \times 10^{-4} < Br(B \to X_s \gamma) < 4.5 \times 10^{-4}$$ 
which is in agreement with CLEO and ALEPH results. Our results are
given in Figs.(\ref{fig:1} - \ref{fig:8}). 

\noindent From our numerical analysis we can conclude :
\begin{enumerate}
\item{} {\bf Branching ratios : } As we can see from
Figure.(\ref{fig:1}) for inclusive mode ($\btoxdll$) that there can be
significant increase in the branching ratio of this decay mode both in
mSUGRA and SUGRA model as compared to SM . This has been stated
earlier on also \cite{Dai:1997vg} in context of $\btoxsll$. As we can
see from Figure(\ref{fig:5}), this pattern (that branching ratio
shows significant increase from SM results) repeats for exclusive mode
($\bdtollg$) again this has earlier on stressed in earlier works
\cite{RaiChoudhury:2002i,Iltan:2000iw}.  
\item{} {\bf FB asymmetries : } As we can see from Figures(\ref{fig:2})
and (\ref{fig:6}) that FB asymmetries also shows fairly large
deviations from SM results both in mSUGRA and SUGRA. Again this point
has been stressed in many earlier works
\cite{Dai:1997vg,Iltan:2000iw,Bobeth:2001jm,RaiChoudhury:2002i}.
\item{}{\bf CP violating partial width asymmetry :}  The effect
scalars on  CP violating asymmetries in exclusive decay modes
$\btopill$ and $\btorholl$ as already been discussed in our earlier
work \cite{Choudhury:2002ab}. There it was shown that the CP violating
partial width asymmetries for both the exclusive modes decrease with
the introduction of scalars in the theory (Higgs here). Here as we can
see from Figure(\ref{fig:3}) that the same trend is present for the
inclusive decay mode $\btoxdll$, but contrastingly, as we can see from
Figure(\ref{fig:7}) the exclusive decay mode $\bdtollg$ doesn't show
up this trend. In fact in this decay mode the CP violating partial
width asymmetry increases with switching on the scalar effects. 
\item{}{\bf CP violation via FB asymmetries :} For estimating this
effect we have introduced $\delta_{FB}$. As we can see from
Figure(\ref{fig:4}) that this parameter follows the trend followed by
$\btopill$ and $\btorholl$ (noted in \cite{Choudhury:2002ab}) , which
is that $\delta_{FB}$ increases with switching on of the scalar
effects as compared to the SM values. But here again as we can see
from Figure(\ref{fig:8}) the trend for $\bdtollg$ is opposite, here in
mSUGRA and SUGRA $\delta_{FB}$ reduces as compared to the SM value. 
\end{enumerate}

Although the branching ratios of both $\btoxdtt$ and $\bdtottg$ are
very low but with upcoming B-factories like LHC-b where more than
$10^{11}$ $B_d$ will be produced, one can hope of observing these
modes. In semi-leptonic decays as far as the branching ratios and FB
asymmetries are concerned, branching ratio tends to increase, and FB
asymmetry tends to decrease with  increasing the scalar effects. This
has been noted in many different decay modes like : $\bstollg$
\cite{Iltan:2000iw,RaiChoudhury:2002i}, $\btopill$ and $\btorholl$
\cite{Choudhury:2002ab,Demir:2002a} , $B_s \to \ell^+ \ell^-$  
\cite{Skiba:1993mg,Bobeth:2001jm,Grossman:1997qj}, $\btoxsll$ 
\cite{Dai:1997vg,Choudhury:1999ze,Bobeth:2001jm} , $B \to (K,K^*)
\ell^+ \ell^-$ \cite{Demir:2002a,Aliev:2000jx}. But as we can see the
CP asymmetries doesn't follow the same trend. For some channels they
decrease and for other they increase. So in brief the measurement of
CP asymmetries although a challenging task, could be very useful for
more information about scalar effects and hence any new physics.


\acknowledgments{
This work was supported under SERC scheme of Department of Science and
Technology, India}  


\appendix


\section{\label{appendix:1} Input parameters and constants } 

\beqa
f(\hat{m}_c) &=& 1 - 8 \hat{m}_c^2 + 8 \hat{m}_c^4 - \hat{m}_c^8 - 24
\hat{m}_c^4 ln (\hat{m}_c) 
\label{appen1:eq:1}     \\
k(\hat{m}_c) &=& 1 - \frac{2 \alpha_s(m_b)}{3 \pi} 
    \Bigg[ \left( \pi^2 - \frac{31}{4} \right)(1 - \hat{m}_c^2) + {3
\over 2} \Bigg]
\label{appen1:eq:2}
\eeqa
The branching ratio of charged current semi-leptonic decay mode
$\btoxcenu$ we are taking to be : 
$$ Br(\btoxcenu) ~=~ 10.4 ~ \% $$ 

The parameters we have used for our numerical analysis are :
$$m_\tau \ = \ 1.77 \ {\rm GeV}$$
$$m_b \ = 4.8 \ {\rm GeV} ~,~ m_c \ = \ 1.4 \ {\rm GeV} ~,~ m_t \ = \
176 \ {\rm GeV} ~,~ \mbd \ = \ 5.26 \ {\rm GeV} $$ 
$$ f_{B_d} \ = \ 1.8 ~,~ \alpha \ = \frac{1}{129} ~,~ \tau_{B} \ = \
1.5 \times 10^{-12} \ {\rm s}$$ 
\noindent Wolfenstein parameters : $$ \rho \ = \ - 0.07 ~,~ \eta \ = \
0.34 ~,~ \lambda \ = \ 0.22 ~,~ A \ = \ 0.84$$ 

\noindent For mSUGRA the parameters we have taken as :
$$m = 200 \ {\rm GeV} ~ , ~ M = 500 \ {\rm GeV} ~,~ A = 0 ~,~ tan\beta
= 45 ~,~ sgn(\mu) = +ve $$ 
 The additional parameter for SUGRA , the pseudoscalar Higgs mass is
taken to be $m_A = 281 \ {\rm GeV}$  


\section{\label{appendix:2} Form factors }

\beqa
A  &=&  \frac{1}{m_{B_d}^2} ~[ \cn G_1(p^2) ~-~  2 \csev
        \frac{m_b}{p^2}G_2(p^2)],                   \nonumber  \\ 
B  &=&  \frac{1}{m_{B_d}^2}~ [ \cn F_1(p^2) ~-~ 2 \csev
        \frac{m_b}{p^2}F_2(p^2)],                   \nonumber  \\ 
C  &=&  \frac{C_{10}}{m_{B_d}^2}~G_1(p^2),           \nonumber  \\
D  &=&  \frac{C_{10}}{m_{B_d}^2}~F_1(p^2).
\label{appen2:eq:1}
\eeqa
In getting above eqns we have used following definitions of
the form factors \cite{Eilam:1995zv} 
\beq
\langle\gamma |~ \bar{d} \gamma_\mu (1 \pm \gamma_5) b ~|B_d \rangle
   =   \frac{e}{m_{B_d}^2}
         \left\{ \epsmualbesig \epsilon_\alpha^* p_\beta q_\sigma 
             G_1(p^2)\mp i [ (\epsilon_\mu^*(pq)-(\epsilon^*p)q_\mu) ]
            F_1(p^2) 
         \right\}                                          
\label{appen2:eq:2}       
\eeq
\beq
\langle\gamma| ~\bar{d} i \sigma_{\mu\nu} p_\nu (1 \pm \gamma_5) b ~
 |B_d\rangle 
  =   \frac{e}{m_{B_d}^2}
        \left\{  \epsmualbesig \epsilon_\alpha^* p_\beta q_\sigma 
           G_2(p^2) \pm i [ (\epsilon_\mu^*(pq)-(\epsilon^*p)q_\mu) ]
           F_2(p^2) 
        \right\}
\label{appen2:eq:3}     
\eeq
another relation we can get by multiplying $p_\mu$ on both the sides
of eqn.(\ref{appen2:eq:3}) :
\beq
\langle\gamma|~ \bar{d} (1 \pm \gamma_5) b ~|B_d\rangle ~=~ 0
\label{appen2:eq:4}
\eeq
Here $\epsilon_\mu$ and $q_\mu$ are the four vector polarization and
momentum of photon respectively.

The defination of the form factors used in above eqns for our
numerical analysis are \cite{Eilam:1995zv} : 
\beqa
G_1(\psq) ~=~ \frac{1}{1 - \psq/5.6^2} ~GeV &,& G_2(\psq) ~=~
\frac{3.74}{1 - \psq/40.5} GeV^2 ,        \nonumber   \\
F_1(\psq) ~=~ \frac{0.8}{1 - \psq/6.5^2} ~ GeV &,& F_2(\psq) ~=~
\frac{0.68}{1 - \psq/30}~GeV^2 . 
\label{appen2:eq:5}
\eeqa

when photon is emitted from lepton lines we use following definations
: 
\beqa
\langle 0|~\bar{d} b ~|B_d\rangle &=& 0 
\label{appen2:eq:6}           \\
\langle 0|~\bar{d}\sigma_{\mu\nu} (1 + \gamma_5) b~|B_d\rangle &=& 0 
\label{appen2:eq:7}                       \\
\langle 0|~ \bar{d} \gamma_\mu \gamma_5 b ~|B_d\rangle &=& -~ i
f_{B_d} P_{B_d\mu}
\label{appen2:eq:8}
\eeqa





\begin{thebibliography}{99}
%


\bibitem{Wolfenstein:1983yz}
  L.~Wolfenstein,
  \prl{51}{1983}{1945}.


\bibitem{RaiChoudhury:1997a}
 S.~Rai Choudhury,
 \prd{56}{1997}{6028} 
 [\hepph{9706313}] , 
\\
 G.~Buchalla, G.~Hiller and G.~Isidori
 \prd{63}{2000}{014015}
 [\hepph{0006136}].


\bibitem{Kruger:1997jk}
 F.~Kruger and L.~M.~Sehgal,
 \prd{56}{1997}{5452}
 [Erratum-ibid.\ D {\bf 60} (1997) 099905]
 [\hepph{9706247}].


\bibitem{Kruger:1996dt}
 F.~Kruger and L.~M.~Sehgal,
 \prd{55}{1997}{2799}
 [\hepph{9608361}].


\bibitem{Skiba:1993mg}
  W.~Skiba and J.~Kalinowski,
  \npb{404}{1993}{3} ,
\\
  H.~E.~Logan and U.~Nierste,
  \npb{586}{2000}{39}
  [\hepph{0004139}].


\bibitem{Dai:1997vg}
  Y.~Dai, C.~Huang and H.~Huang,
  \plb{390}{1997}{257}
  [\hepph{9607389}] ;
\\
  Z.~Xiong and J.~M.~Yang,
  \npb{628}{2002}{193}
  [\hepph{0105260}] ;
\\
  C.~Huang, W.~Liao and Q.~Yan,
  \prd{59}{1999}{011701}
  [\hepph{9803460}] .


\bibitem{Choudhury:1999ze}
  S.~R.~Choudhury and N.~Gaur,
  \plb{451}{1999}{86}
  [\hepph{9810307}].
\\
  P.~H.~Chakowski and L.~Slawianowska,
  \prd{63}{2001}{054012} ,
  [\hepph{0008046}]. 
\\
  A.~Dedes, H.~K.~Dreiner and U.~Nierste,
  \prl{87}{2001}{251804} ,
  [\hepph{0108037}] . 
\\
  K.~S.~Babu and C.~Kolda,
  \prl{84}{2000}{228}
  [\hepph{9909476}];
\\
  R.~Arnowitt, B.~Dutta, T.~Kamon and M.~Tanaka,
  \hepph{0203069} .


\bibitem{Ali:1997vt}
  A.~Ali,
  \hepph{9709507} ; 
\\
  G.~Buchalla, A.~J.~Buras and M.~E.~Lautenbacher,
  \rmp{68}{1996}{1125} [\hepph{9512380}].

\bibitem{Iltan:2000iw}
  E.~O.~Iltan and G.~Turan,
  \prd{61}{2000}{034010}
  [\hepph{9906502}] ,
\\
  G.~Erkol and G.~Turan , 
  \hepph{0112115}

\bibitem{Choudhury:2002ab}
 S.~Rai Choudhury and N.~Gaur ,
 \hepph{0206128}






\bibitem{Bobeth:2001jm}
  C.~Bobeth, A.~J.~Buras, F.~Kruger and J.~Urban,
  \npb{630}{2002}{87} [\hepph{0112305}] ; 


\bibitem{RaiChoudhury:1999qb}
  S.~Rai Choudhury, A.~Gupta and N.~Gaur,
  \prd{60}{1999}{115004}
  [\hepph{9902355}] ;
\\
  S.~R.~Choudhury, N.~Gaur and A.~Gupta,
  \plb{482}{2000}{383}
  [\hepph{9909258}].


\bibitem{RaiChoudhury:2002i}
 S.~Rai Choudhury, N.~Gaur and N.~Mahajan , 
 to be published in Phys. Rev. D [\hepph{0203041}] , 
\\
 S.~Rai Choudhury and N. Gaur
 \hepph{0205076}. 


\bibitem{Demir:2002a}
 D.~A.~Demir, K.~A.~Oliev and M.~B.~Voloshin
 \hepph{0204119}.
\\
 G.~Erkol and G.~Turan
 \hepph{0201055} . 




\bibitem{goto1} 
  T.~Goto, Y.~Okada, Y.~Shimizu and M.~Tanaka,
  \prd{55}{1997}{4273}
  [\hepph{9609512}],
\\
  T.~Goto, Y.~Okada and Y.~Shimizu,
  \prd{58}{1998}{094006}
  [\hepph{9804294}].
\\
  J.~R.~Ellis, K.~A.~Olive and Y.~Santoso
  \hepph{0204192}


\bibitem{Cho:1996we}
  P.~L.~Cho, M.~Misiak and D.~Wyler,
  \prd{54}{1996}{3329}
  [\hepph{9601360}], 
\\
  J.~L.~Hewett and J.~D.~Wells,
  \prd{55}{1997}{5549}
  [\hepph{9610323}].


\bibitem{Du:1995ez}
 D.~S.~Du and M.~Z.~Yang,
 \prd{54}{1996}{882}
 [\hepph{9510267}],
\\
 T.~M.~Aliev, D.~A.~Demir, E.~Iltan and N.~K.~Pak,
 \prd{54}{1996}{851}
 [\hepph{9511352}].







\bibitem{Aliev:2000jx}
  T.~M.~Aliev, M.~K.~Cakmak and M.~Savci,
   \npb{607}{2001}{305}
  [\hepph{0009133}] ,
\\
  T.~M.~Aliev, M.~K.~Cakmak, A.~Ozpineci and M.~Savci,
  \prd{64}{2001}{055007}
  [\hepph{0103039}]; 
\\
  Q.~S.~Yan, C.~S.~Huang, W.~Liao, S.~H.~Zhu ,
  \prd{62}{2000}{094023} ,
  [\hepph{0004262}] . 
\\
  C.~Bobeth, T.~Ewerth, F.~Kruger and J.~Urban,
  \prd{64}{2001}{074014} 
  [\hepph{0104284}].



\bibitem{Kruger:1996cv}
  F.~Kr\"{u}ger and L.~M.~Sehgal,
  \plb{380}{1996}{199}
  [\hepph{9603237}] ; 



\bibitem{Grossman:1997qj}
  Y.~Grossman, Z.~Ligeti and E.~Nardi,
  \prd{55}{1997}{2768}
  [\hepph{9607473}].


\bibitem{Grinstein:1989me}
  B.~Grinstein, M.~J.~Savage and M.~B.~Wise,
  \npb{319}{1999}{271} ;
\\
  A.~J.~Buras and M.~M\"{u}nz,
  \prd{52}{1995}{186}
  [\hepph{9501281}].


\bibitem{Long-Distance}
  A.~Ali, T.~Mannel and T.~Morozumi,
  \plb{273}{1991}{505} . 
\\
  C.~S.~Lim, T.~Morozumi and A.~I.~Sanda,
  \plb{218}{343}{1989} .
\\
  N.~G.~Deshpande, J.~Trampetic and K.~Panose,
  \prd{39}{1461}{1989} .
\\
  P.~J.~O'Donnell and H.~K.~Tung,
  Phys.\ Rev.\ D {\bf 43}, 2067 (1991) .


\bibitem{expbsg}
 B Physics at Tevatron : Run II \& Beyond ~,~ K. Anikeev \etal
 \hepph{0201071} ;   
\\
 CLEO collaboration, T.E.Coan \etal 
 \prl{84}{2000}{5283}
 [\hepex{9908022}] ; 
\\
 ALEPH Collaboration, R.Barate \etal
 \plb{429}{1998}{169} .



\bibitem{Aliev:1997ud}
  T.~M.~Aliev, A.\"{O}zpineci and M.~Savci,
  \prd{55}{1997}{7059} ,  [\hepph{9611393}] ; 
\\
  T.~M.~Aliev, N.~K.~Pak and M.~Savci,
  \plb{424}{1998}{175} , [\hepph{9710304}].



\bibitem{Eilam:1995zv}
  G.~Eilam, I.~Halperin and R.~R.~Mendel,
  \plb{361}{1995}{137} ,  [\hepph{9506264}].
\\
  T.~M.~Aliev, A.\"{O}zpineci and M.~Savci,
  \prd{55}{1997}{7059} ,  [\hepph{9611393}] ; 


\end{thebibliography}
\end{document}